\documentclass[pdflatex,sn-mathphys]{sn-jnl}% Math and Physical Sciences Reference Style

\jyear{2023}%

\theoremstyle{thmstyleone}%
\theoremstyle{thmstyletwo}%

\theoremstyle{thmstylethree}%

\begin{document}

\title[Applicability of Trust Management Algorithm in C2C services]{Applicability of Trust Management Algorithm in C2C services} 

\author*[1]{\fnm{Ryohei} \sur{Suzuki}}\email{ryohei.suzuki.kv[at]hco.ntt.co.jp}
\author[1]{\fnm{Iifan} \sur{Tyou}}\email{iifan.tyou.tm[at]hco.ntt.co.jp}
\author[1]{\fnm{Shigenori} \sur{Ohashi}}\email{shigenori.ohashi.ur[at]hco.ntt.co.jp}
\author[2]{\fnm{Kazutoshi} \sur{Sasahara}}\email{sasahara.k.aa[at]m.titech.ac.jp}

\affil*[1]{\orgname{NTT Social Informatics Laboratories}, \orgaddress{\street{3-9-11 Midori-cho, Musashino-shi}, \city{Tokyo}, \postcode{180-8585},\country{Japan}}}
\affil[2]{\orgdiv{School of Environment and Society}, \orgname{Tokyo Institute of Technology}, \orgaddress{\street{3-3-6, Shibaura, Minato-ku}, \city{Tokyo}, \postcode{105-0023}, \country{Japan}}}

\abstract{
The emergence of Consumer-to-Consumer (C2C) platforms has allowed consumers to buy and sell goods directly, but it has also created problems, such as commodity fraud and fake reviews. Trust Management Algorithms (TMAs) are expected to be a countermeasure to detect fraudulent users. However, it is unknown whether TMAs are as effective as reported as they are designed for Peer-to-Peer (P2P) communications between devices on a network. Here we examine the applicability of `EigenTrust', a representative TMA, for the use case of C2C services using an agent-based model. First, we defined the transaction process in C2C services, assumed six types of fraudulent transactions, and then analysed the dynamics of EigenTrust in C2C systems through simulations. We found that EigenTrust could correctly estimate low trust scores for two types of simple frauds. Furthermore, we found the oscillation of trust scores for two types of advanced frauds, which previous research did not address. This suggests that by detecting such oscillations, EigenTrust may be able to detect some (but not all) advanced frauds. Our study helps increase the trustworthiness of transactions in C2C services and provides insights into further technological development for consumer services.
}
%%=======================
\keywords{
Consumer to Consumer service, Trust Management Algorithm, Trust%Reputation 
}

\maketitle

\section{Introduction}\label{introduction}
The Internet enables us to acquire various information and complete contracts online. 
In particular, digital platform technology has given rise to a series of Consumer-to-Consumer (C2C) services in which users transact directly with each other, and these business ecosystems are creating new value. 
For example, e-commerce services have become widespread, and many people now use significant online marketplaces. 
For example, Amazon allows stores to sell to users online~\cite{amazon}. 
Additionally, C2C e-commerce services, such as eBay, Mercari, and Taobao, rapidly gained popularity by allowing users to both sell and buy directly ~\cite{ebay,mercari,taobao}.

While C2C platforms have increased user convenience, the credibility of transactions has become an issue, with fraudulent products being listed on the market and the reputations of counterparties being intentionally lowered. 
Such fraudulent actions are easy to do because consumers can purchase goods impersonally and even consumers can participate as providers~\cite{TrustGuard,security_of_reputation,trust_and_reputation_system}.

C2C platforms have been working to detect fraudulent users to improve the trustworthiness of transactions. One such attempt is to use a Trust Management Algorithm (TMA)~\cite{EigenTrust,Risk-analysis}. TMAs have been developed to calculate the trust scores of distributed nodes in Peer-to-Peer (P2P) communication networks rather than C2C services. In TMAs, a rating is performed between nodes as each node communicates with the other, and the accumulated rating information is used to quantify how trustworthy each node is. For example, EigenTrust is a typical TMA that calculates each user's rating values by recursively aggregating the rating values of the entire network. PeerTrust is another type of TMA that was designed to calculate trust scores for a specific node by considering the similarity between raters and referents of their ratings~\cite{peertrust}.

As TMA is a technology considered in the ideal state in P2P communications, it does not consider preconditions for users and transaction procedures among users in C2C services. Thus, it is unclear how effective TMA is at detecting malicious users and groups who commit fraud in actual C2C transactions. For the growing business ecosystem of C2C platforms, a pressing issue is to visualise the trustworthiness of user transactions and prevent fraudulent activities by malicious users.

In this study, we investigated the applicability of EigenTrust for the use case of C2C services using an agent-based model. In particular, we introduced realistic settings, such as transaction processes between users and fraudulent actions of users, into the model. We analysed the simulated behaviours of EigenTrust to different types of fraudulent transactions to identify its possibilities and limitations in the context of C2S services.

%--------------------------------------------------------------------------------------------------------------------
\section{Methods}\label{Methods}
In this study, we assume a use case of buying and selling in C2C services and set up conditions such as transaction procedures between users. We developed a simulator of an agent-based model that performs transactions on the basis of the set conditions and conducted simulations to see if EigenTrust can detect users who are committing fraudulent acts (`attackers').

\subsection{EigenTrust}
TMA calculates the trust scores of each node on the basis of the rating information of all nodes in a network. Again, EigenTrust is a type of TMA developed for P2P file-sharing networks to reduce the number of unauthorised file downloads. EigenTrust calculates a unique, system-wide, global trust score for each node on the basis of ratings of the success or failure of all nodes' communications.
%%---
EigenTrust calculates the trust score as follows. The rating value $s_{ai}$ from node $a$ to node $i$ is normaliszed as $c_{ai}=\frac{\max (s_{ai},0)}{\sum_{i} \max (s_{ai},0)}$ and similarly the rating value $s_{ij}$ of node $i$ is normaliszed as $c_{ij}=\frac{\max (s_{ij},0)}{\sum_{j} \max (s_{ij},0)}$. The trust score of the target $j$ concerning rater $a$ is computed by $\sum_{i} c_{ai} c_{ij} $ (the sum is over all nodes $a$ and adjacent nodes $i$ that refer to the trust score). 
This trust score calculation indicates that from the column vector $c_a$, whose elements are the ratings $c_{ai}$ of the acquaintances of the referee $a$, the trust score can be calculated by using the matrix operation $(C^T c_a)$ (where $C$ is a matrix whose elements are $c_{ij}$). From the $c_{ij}$ matrix, we can compute the trust score of a subject reflecting the opinion of an acquaintance (not a direct acquaintance) of $a$. By repeating this procedure $n$ times, we can obtain a trust score vector $t=(C^T)^n c_{a}$ that converges from nodes with a track record of transactions to reflect further the opinions of nodes with a track record of transactions.

When applying EigenTrust to C2C services, we need to consider the frequency of transaction occurrences and the variability of communication targets specific to C2C services, which may differ from those of P2P communications.
Therefore, it is not apparent whether EigenTrust can detect fraudulent activities.

\subsection{Threat Models}\label{threat_model_def}
We consider different types of fraudulent user activities to evaluate the applicability of EigenTrust in the context of C2C services. 
To this end, we used six types of threat models defined in the previous studies~\cite{EigenTrust,grouptrust,servicetrust++,Risk-analysis}. 
Simulating these attacks allows us to compare the behaviours of EigenTrust under the same threat models in C2C and P2P contexts.

First, we define two types of `attack' from the viewpoint of malicious users. These are constituents of the threat models that we will explain next.

\begin{itemize}
\item \textbf{Service attack}: 
An attacker sells a defective product to a user (buyer).

\item \textbf{Rating attack}: 
An attacker as a buyer gives a bad rating to a user (seller).
\end{itemize}

There is mutual evaluation on C2C platforms (as well as P2P). In the case of the service attack, the buyer receives low-quality goods and therefore evaluates the attacker lower (i.e. lower trust score). In the case of the rating attack, the seller receives an unfair rating irrespective of the quality of goods and therefore evaluates the attacker lower. 
When `collusion' among attackers is assumed, they give unfairly high ratings to each other (explained later).

Second, the threat models (A--F) are defined by the probability and the timing of the two types of attack and the division of roles among malicious users or attackers (Table~\ref{tab:threat_model}). These models were introduced in previous studies~\cite{EigenTrust,grouptrust,servicetrust++,Risk-analysis}. 
The differences between models are illustrated in Fig.~\ref{threat_model_def}, where the complexity of the model varies from left to right. 

%-------------------------------------------------------------------
\renewcommand{\thefootnote}{\fnsymbol{footnote}}	
\begin{table}[h]
\begin{center}
\begin{minipage}{\textwidth}%{174pt}
\caption{Differences between threat models}\label{tab:threat_model}%
\begin{tabular}{@{}cll|c@{}}
\toprule
Threat Model   & Service Attack Prob.   & Rating Attack Prob. & EigenTrust Coverage(\cite{Risk-analysis})\\
\midrule
Model A    & 1\footnotemark[1]       & 1\footnotemark[1]     & +\\
Model B    & 1                       & 1                     & +\\
Model C    & 1-$c$                     & 0                     & -\\
Model D    & [0,1]\footnotemark[2]   & 1                     & -\\
Model E    & 1-$c$                     & 1-$e$                   & -\\
Model F    & [0,1]\footnotemark[2]   & 1-$f$                   & -\\
\botrule
\end{tabular}
\footnotetext{$c$, $e$, and $f$ are parameters for the probability of camouflage without attacking.}
\footnotetext{"+" indicates EigenTrust can detect threat model, oppositely "-" indicates EigenTrust cannot.}
\footnotetext[*]{Independent attacks (without collusion)}
\footnotetext[\dag]{In addition to regular attackers, spies are included.}
\end{minipage}
\end{center}
\end{table}
%-------------------------------------------------------------------

% Fig. 1
\begin{figure}[h]
    \centering
    \includegraphics[width=\textwidth]{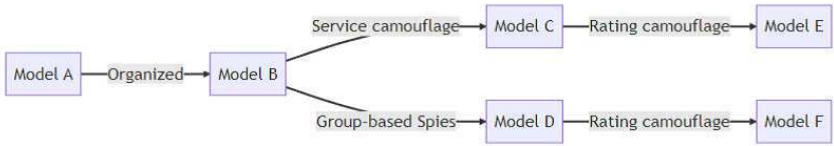}
    \caption{Difference between Threat Models}
    \label{threat_models_diff}
\end{figure}

\begin{enumerate}
\item \textbf{Model A (Independently Mischievous)}:
This type of attacker performs both service and rating attacks and acts independently. 
This is the simplest threat model.

\item \textbf{Model B (Collectively Mischievous)}:
Model B follows Model A, but attackers collude to perform service and rating attacks. 
Here `collude' is used to indicate collective fraud by a group of attackers. 
When colluding attackers buy and sell with each other, the buyer will always give a high rating, thereby leading to an unjustified increase in the trust scores of the selling attackers. 
When colluding attackers buy and sell with others, they always give sellers a low rating when buying and give buyers a defective product when selling.

\item \textbf{Model C (Camouflage Collective)}:
Model C follows Model B but with a stochastic service attack, which is referred to as `camouflage'. 
The attackers sometimes provide appropriate services, thereby pretending to be non-attackers. Colluding attacks are also applied in this setting.

\item \textbf{Model D (Group-based Spies)}:
Model D also follows Model B, but there are two groups: one group consisting of attackers for the service attack only and the other group consisting of attackers for the rating attack only (called `spies'). 
They also conduct colluding attacks, but there are several differences due to the division of labor between the two groups. 
That is, spies sell proper products when they are on the sales side, while they always give a low rating when they are on the purchase side. 
This passive attack makes it harder for others to find the specific group responsible for the service attacks. 

\item \textbf{Model E (Camouflage Collective with Honest Rating)}:
Model E follows Model C except for the rating attack, i.e. attackers perform a stochastic rating attack, in addition to a stochastic service attack. 
They sometimes provide appropriate ratings (as buyers) in addition to selling proper products (as sellers), thereby pretending to be non-attackers (camouflaging). Colluding attacks are also applied in this setting.

\item \textbf{Model F (Group-based Spies with Honest Rating)}:
Model F follows Model D but with a stochastic rating attack. Similar to Model E, this stochastic rating allows attackers to pretend to be non-attackers. In addition, similar to Model D, there are two groups: spies and others. 
Colluding attacks are also applied in this setting. 

\end{enumerate}

As shown in Table~\ref{tab:threat_model}, the previous studies have demonstrated that EigenTrust can detect two simple frauds (i.e. Models A and B) in the case of P2P communications.

\subsection{Transaction}\label{transaction}
EigenTrust assumes a P2P communication procedure ~\cite{gnutella_protocol} in which a node requests to download a file and a satisfactory evaluation is given if the downloaded file is correct. The evaluation is one way. 
In C2C services, on the other hand, users conduct transactions directly with each other and mutual evaluations between the seller and the buyer. 
We therefore assume the following procedures for evaluation on a C2C service.

Let User A be the service provider who sells the goods and User B be the beneficiary of the service who buys the goods.
\begin{description}\label{sequence}
   \item[Providing service:]\mbox{}\\
        User B searches for a provider from whom to purchase a product. Selection is made in accordance with the rules described below. 
        User A then offers the product to User B. 
   \item[Rating:]\mbox{}\\
        User B receives the goods and rates whether the transaction was satisfactory.
        The rating is handled as a binary True/False value of whether the transaction was satisfactory or not.
   \item[Satisfaction rating:]\mbox{}\\
    User A rates whether User B's response was satisfactory. The rating is handled as a True/False binary value of satisfactory or unsatisfactory. The purpose is to reduce the trust score of User B when User A, who provides the service, unilaterally receives an unfair rating from User B.
\end{description}

User B is more likely to buy a product from a user with whom he already has a track record of transactions than from a party with whom he has no track record. In P2P communication, a node is selected on the basis of the content distribution model~\cite{p2psharing,p2pdist}. 
Therefore, we decided to establish the following rules for selecting a buying counterparty.

\begin{description}\label{choise}
   \item[Choice of whether the transaction is from a proven partner:]\mbox{}\\
        A node whose edge is connected to User B is considered as a user with a proven transaction history. In this paper, with a probability of 50\%, we choose the candidate of the purchase destination as the counterparty with no transaction track record or the counterparty with a transaction track record.
   \item[Determining if the counterparty is a trusted counterparty:]\mbox{}\\
        A counterparty is selected on the basis of the trust score of each candidate user from the candidate buyers selected in the previous step. In this paper, we assume that the quality of service may temporarily decline even for normal users and select a partner with a probability weighted by the trust score.
\end{description}

\subsection{Network}\label{network}

In C2C services, users can be either providers, such as those who mainly sell products, or beneficiaries, such as those who mainly buy products. 
Such characteristics in transactions may shape a different network structure from that of P2P communications. 
It is assumed that users have made at least one purchase and that there are no isolated users. We generate a network using `random\_k\_out\_graph,' which generates a network without a degree greater than or equal to $k$~\cite{randomkout,koutgraph}. 
This network is realistic in the sense that it has a skewed out degree distribution with hubs, which is often observed in social networks. 
The initial state of the network is a directed graph generated by the random\_k\_out\_graph function of Python's NetworkX library.

\begin{figure}
    \centering
    \includegraphics[width=0.9\textwidth]{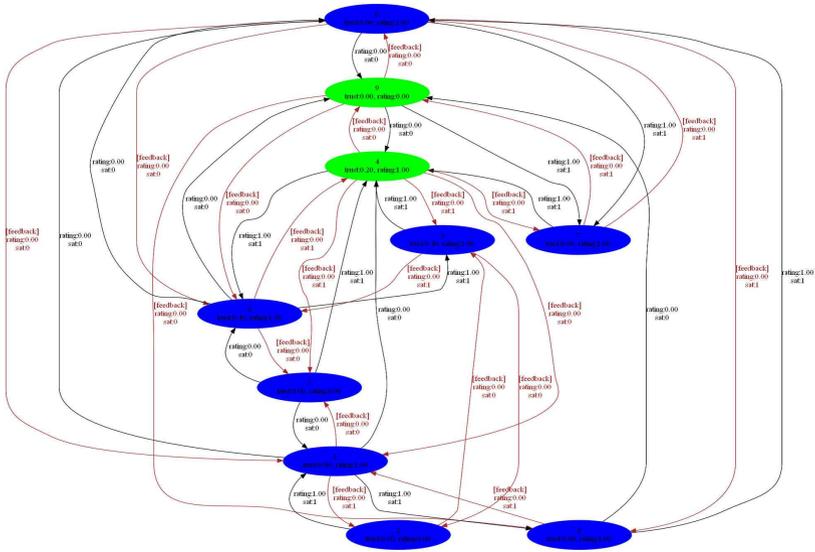}
    \caption{An example of random\_k\_out\_graph}
    \label{fig:example_nw}
\end{figure}

Figure~\ref{fig:example_nw} shows an example graph generated by the random\_k\_out\_graph function. Here the number of nodes is set to 10 for visualisation purposes only. Blue nodes represent normal users, and green nodes represent attackers. In this study, the attackers are always placed next to each other. We set it to 100 for simulations (Table~\ref{params}).

% Fig. 2

\subsection{Behavioural Rules of Attackers}\label{attackers}
An attacker performs the actions specified in \ref{threat_model_def} from the moment of joining the network.
In C2C services, the transaction history of the counterparty is often referenced. This is because if the purchase of a product is unsatisfactory, a financial loss is incurred.
When examining transaction partners, many users may hesitate to buy a product from a user who has no transaction history at all.
Therefore, we assume that C2C attackers are subject to an incubation period.
For this period, attackers behave like normal users and earn trust scores the same as normal users.

%--------------------------------------------------
% Table 2
\renewcommand{\thefootnote}{\alph{footnote}}	
\begin{table*}[h]
\begin{center}
\begin{minipage}{\textwidth}
\caption{Simulation Parameters}\label{params}
\begin{tabular*}{\textwidth}{@{\extracolsep{\fill}}lccc@{\extracolsep{\fill}}}
    \toprule
                  & Configurable parameter                 & Simulation value\\
    \midrule
    Transaction     &   Bidirectional rating                      &  True \\
                    &   Overall Transaction Count                      &  100 times \\
   \addlinespace[5mm]   % --- booktabs: 

    Network         &   Graph type       &  Generated by random\_k\_out\_graph \\
                    &    Total nodes                    & 100 people \\
                    &    Pre-trusted nodes                 & 32 people\% \\
   \addlinespace[5mm]   % --- booktabs: 

    Attacker        &   Attacker ratio                      & 10\% of total nodes \\
                    &   Spy ratio                    &  50\% of attacker nodes\footnotemark[1]  \\
                    &   Attack probability($c$,$e$,$f$)                   &  50\% \footnotemark[2]  \\
                    &   Incubation period               &  50 times    \\
   \botrule
\end{tabular*}
\footnotetext[1]{Only threat models that include spies are considered.}
\footnotetext[2]{The probabilities of $c$, $e$, and $f$ can be set independently. In this paper, they set these uniformly.}
\end{minipage}
\end{center}
\end{table*}
%--------------------------------------------------

\subsection{Simulation}\label{execution}
We developed an agent-based model simulator to simulate C2C services.\footnote{The simulator program is available from the corresponding author at a reasonable request.}
The important parameters used for simulations are summarised in Table~\ref{params}.

First, `Transaction' includes a parameter set related to transactions.
It is possible to set whether both sales and purchases are evaluated in transactions, and the overall number of transactions to be simulated.
In this implementation, the number of transactions is treated as an elapsed time.
Second, `Network' is a configuration item for the assumed user network structure.
The graph type, the total number of nodes, and the percentage of nodes that can be trusted can be set in advance.
These items characterise the target purchasing service and can be set to specify the type of connections assumed between users.
Third, `Attacker' is a configuration item for attackers.
The percentage of attackers, the percentage of spies, the attack probability, and the attack initiation timing can be set, which are related to the degree of attack in the C2C service.

%With these settings, the simulation was performed in the environment shown in Table\ref{tab:environment}.
We assume that users can refer to not only the current value of the trust score of the counterparty but also the past scores and output the change in the trust score over time, thanks to the function of visualisation and database provided by a C2C platform.

%------------------------------------------------
% Fig.3-8
\begin{figure}[htbp]
\centering
    \begin{tabular}{cc}
      \begin{minipage}[t]{0.4\hsize}
        \centering
        \includegraphics[keepaspectratio, scale=0.27]{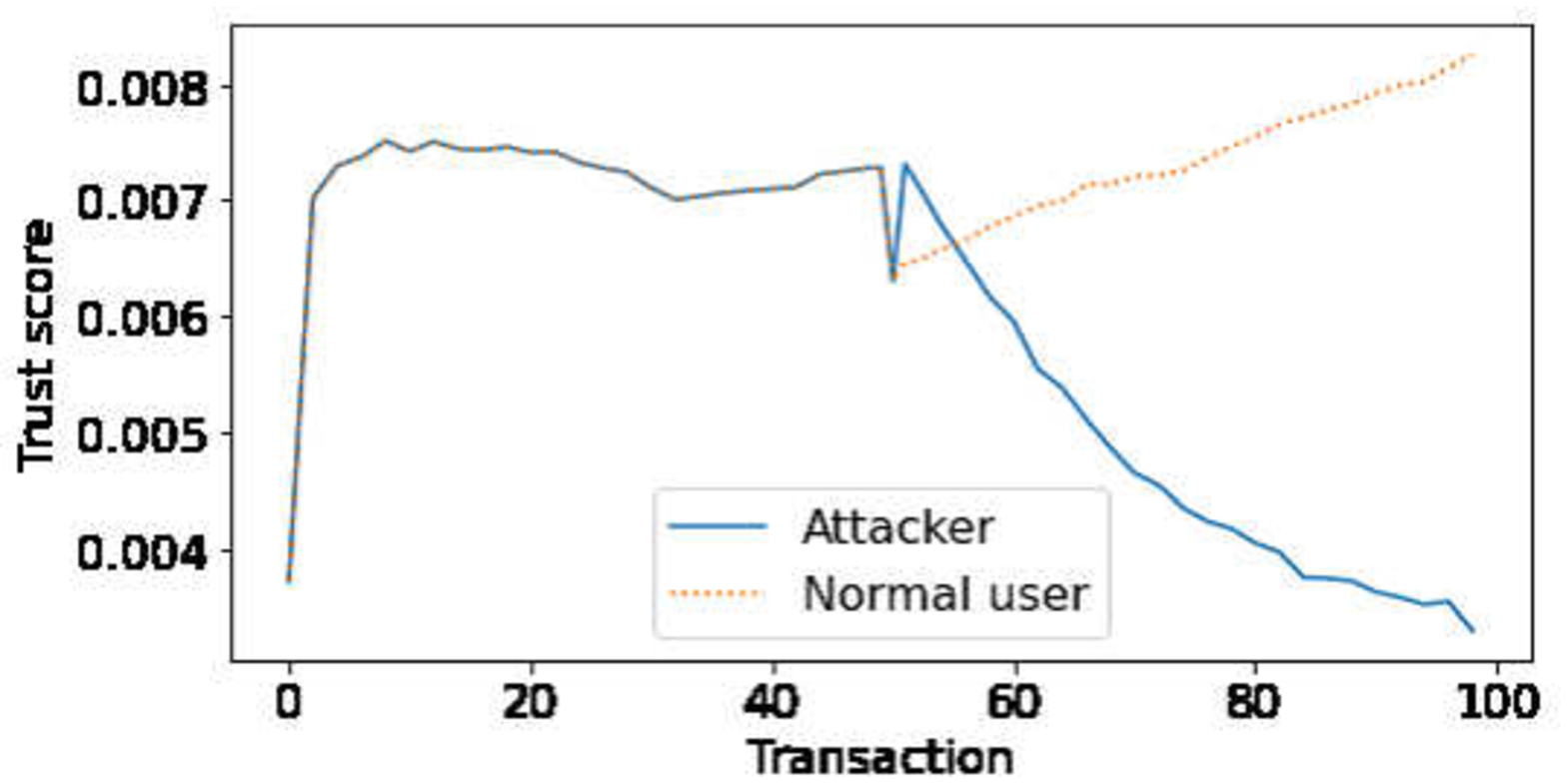}
        \caption{Model A}\label{res_a}
      \end{minipage} &
      
      \begin{minipage}[t]{0.4\hsize}
        \centering
        \includegraphics[keepaspectratio, scale=0.27]{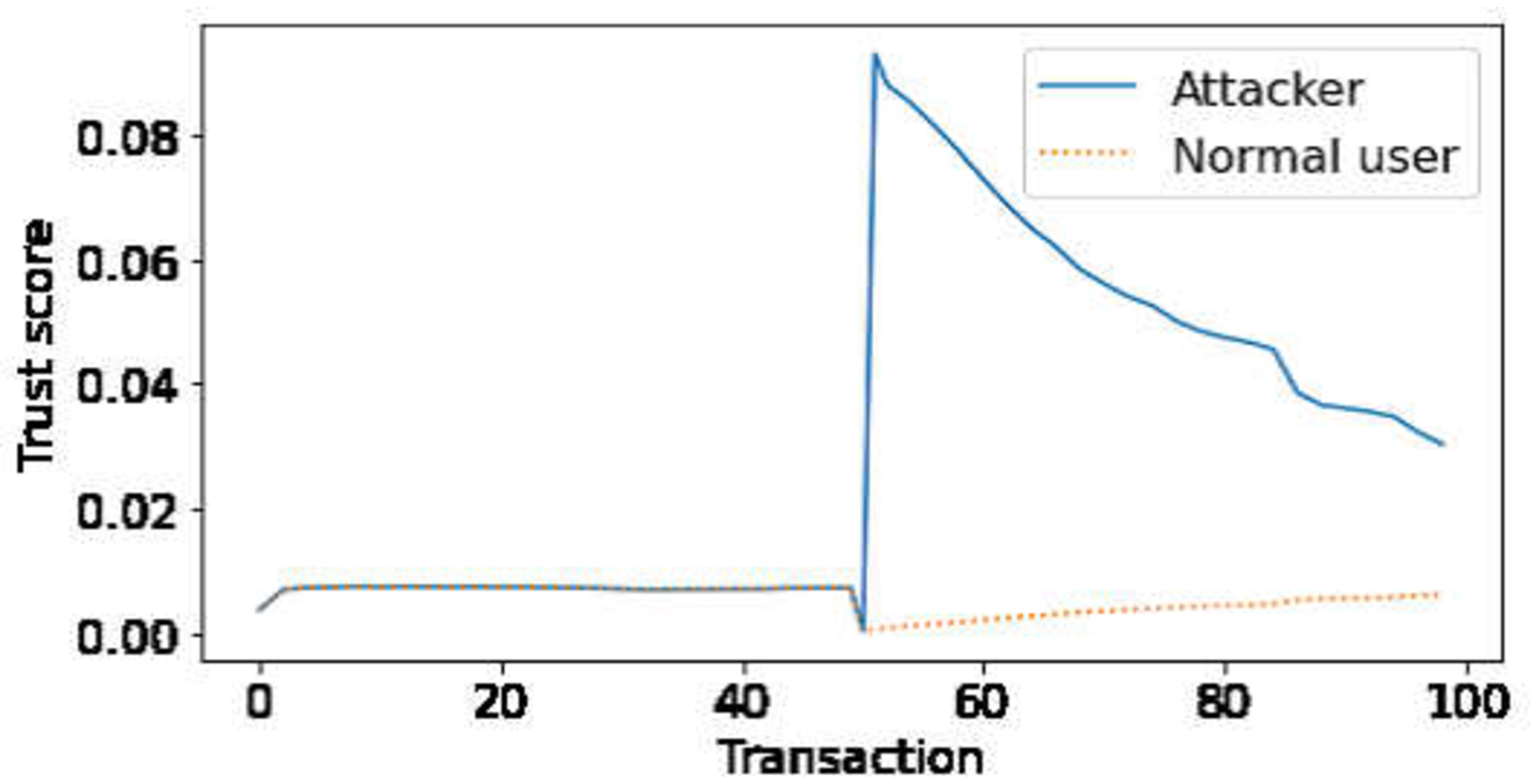}
        \caption{Model B}\label{res_b}
      \end{minipage} \\
   
      \begin{minipage}[t]{0.4\hsize}
        \centering
        \includegraphics[keepaspectratio, scale=0.27]{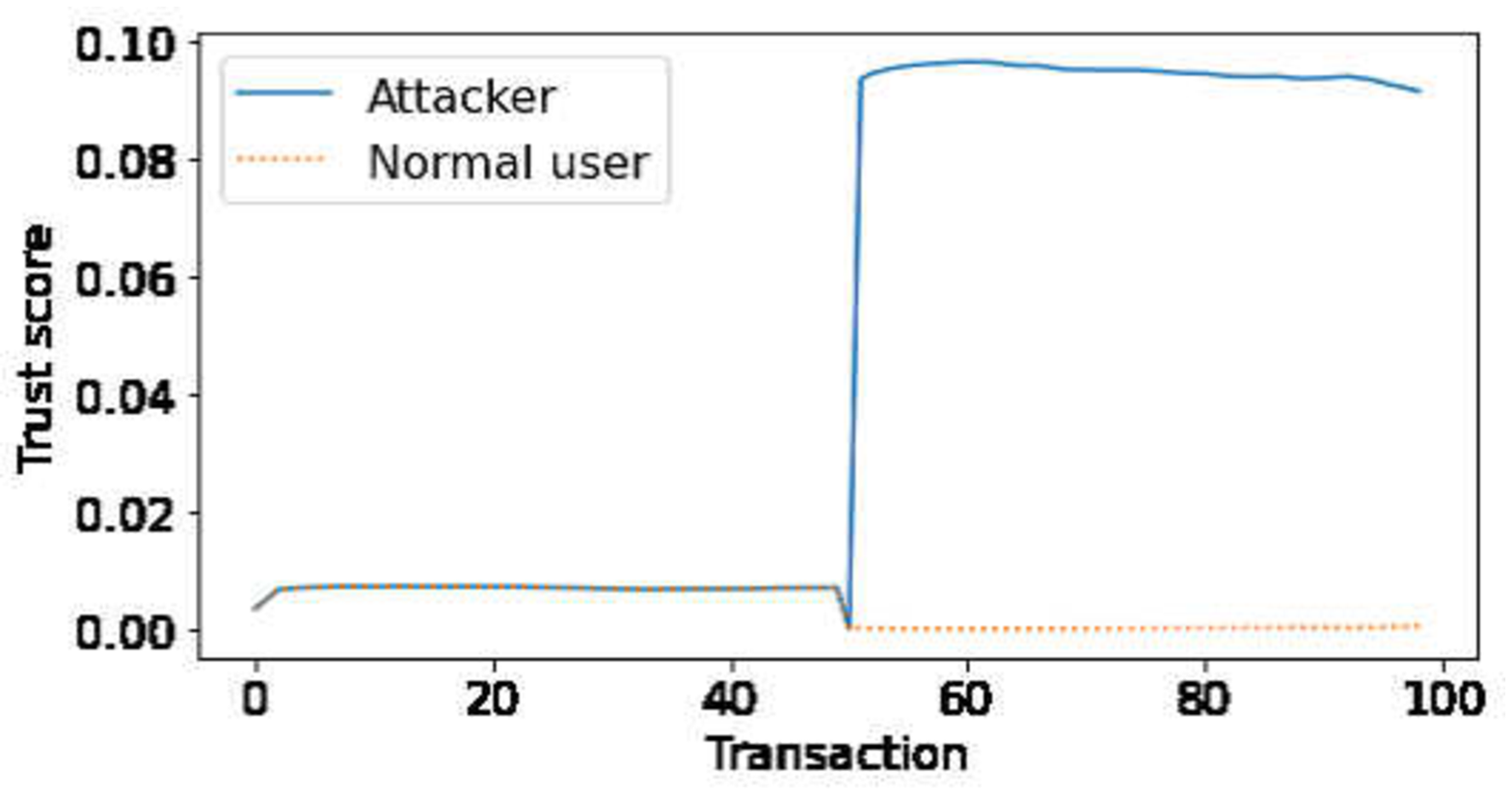}
        \caption{Model C}\label{res_c}
      \end{minipage} &
      
      \begin{minipage}[t]{0.4\hsize}
        \centering
        \includegraphics[keepaspectratio, scale=0.27]{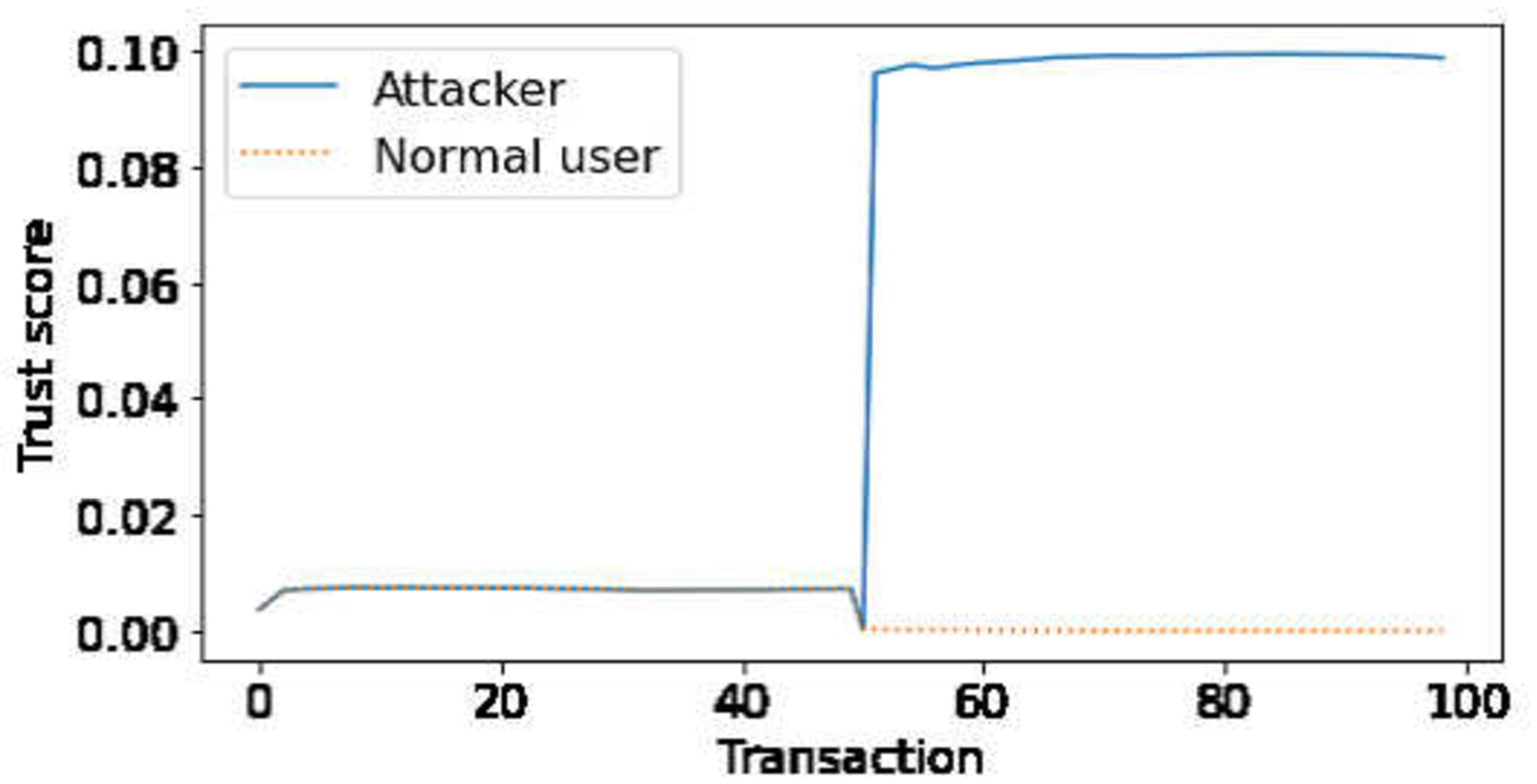}
        \caption{Model D}\label{res_d}
      \end{minipage} \\
      
      \begin{minipage}[t]{0.4\hsize}
        \centering
        \includegraphics[keepaspectratio, scale=0.27]{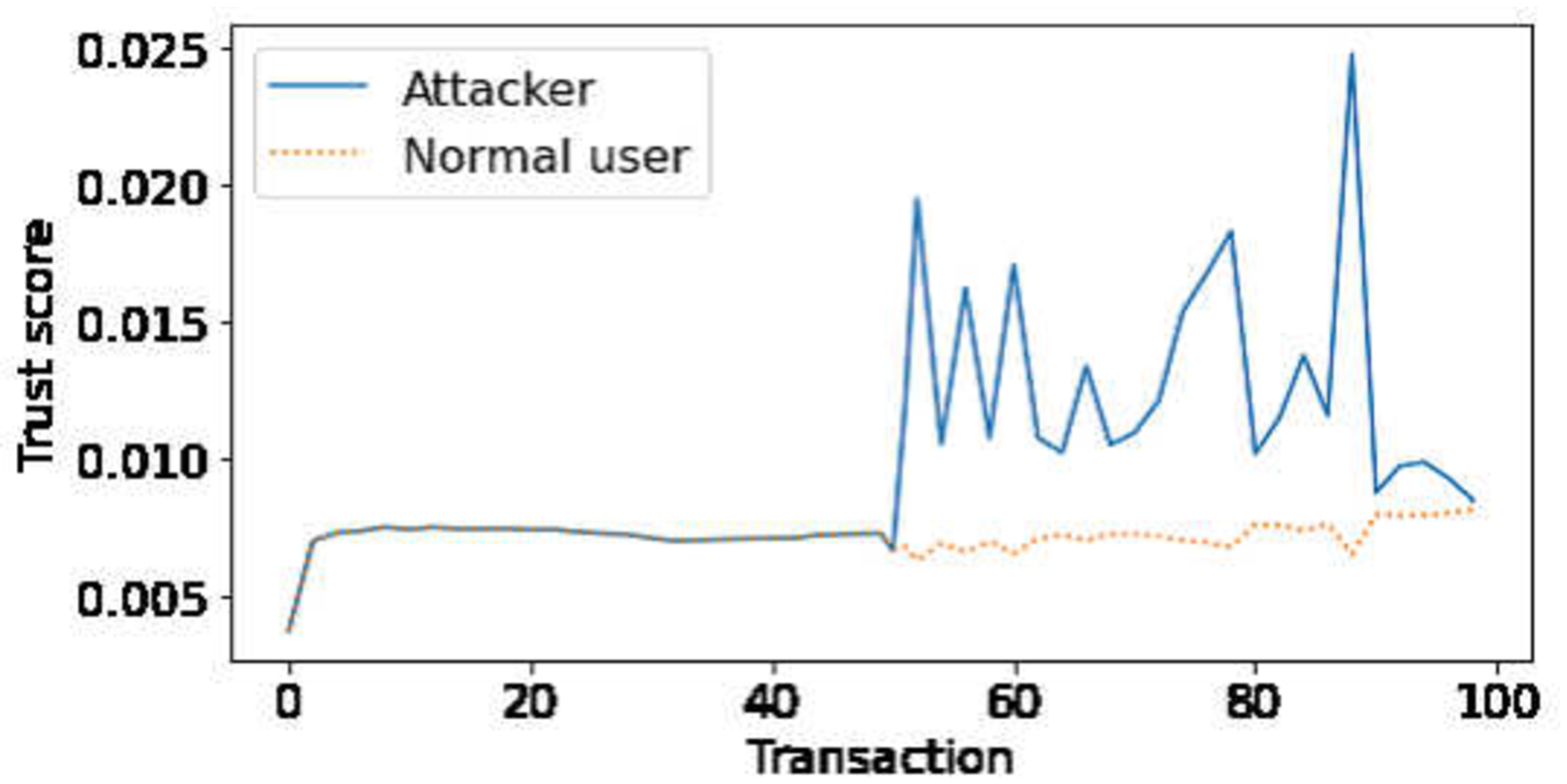}
        \caption{Model E}\label{res_e}
      \end{minipage} &
      
      \begin{minipage}[t]{0.4\hsize}
        \centering
        \includegraphics[keepaspectratio, scale=0.27]{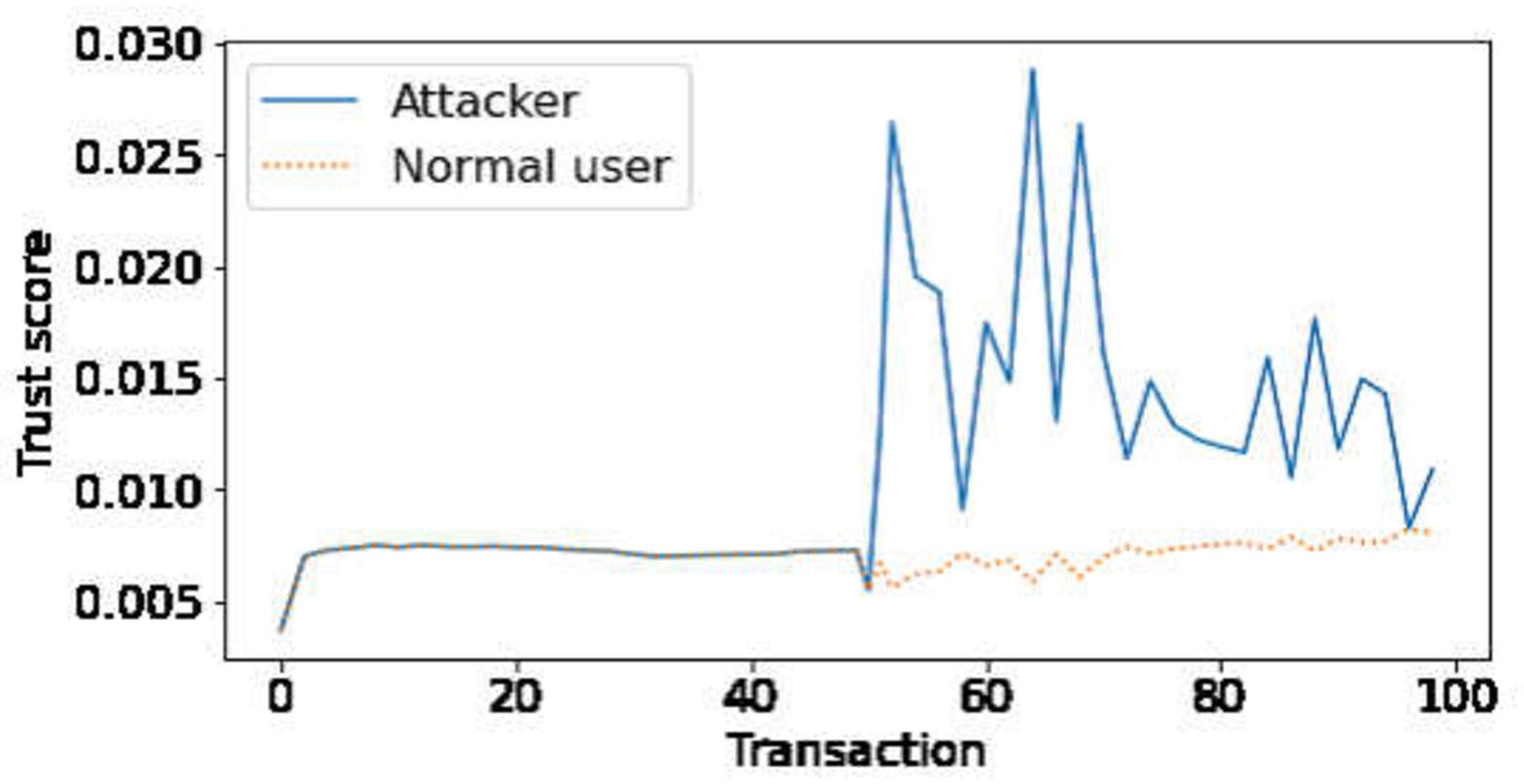}
        \caption{Model F}\label{res_f}
      \end{minipage} 
      
    \end{tabular}
    %\caption{Dynamics of mean trust scores}
\end{figure}
 
%------------------------------------------------

%--------------------------------------------------------------------------------------------------------------------
\section{Results}\label{results} %結果のみを記述
We simulated the dynamics of EigenTrust in six types of frauds (threat models) using the parameter set in Table 2. 
Figs.~\ref{res_a}--\ref{res_f} show temporal changes in the mean trust scores of attackers and normal users in each threat model.
The horizontal axis is the elapsed time (transaction count).

From the 0th to 49th transactions, it is latency time, and the trust score is computed the same as that of a normal user. Then, after the 50th transaction, the attackers start attacking.
The vertical axis is the trust score calculated by EigenTrust, which is the mean trust scores of the attackers.

In Model A (Fig.~\ref{res_a}), the trust score calculated by EigenTrust monotonically decreases as time passes from the start of the attack.
In Model B (Fig.~\ref{res_b}), the trust score increases immediately after the attack starts. After that, the trust score decreases with time.
These results indicate that EigenTrust can properly detect fraudulent activities by these simple attacks.

Models C and D (Fig.~\ref{res_c} and \ref{res_d}) show an abrupt increase and then plateau over time.
This indicates that EigenTrust failed to detect fraudulent activities by these complex attacks.

We found a novel pattern in Models E and F (Fig.~\ref{res_e} and \ref{res_f}). These figures show that the trust score did not converge to a constant and instead oscillated over time.
This pattern was not discussed either C2C or P2P contexts before.
We will discuss mechanistic and technological insights into this trust oscillation phenomenon in Discussion.

% It was confirmed that the trust scores for Threat Models C and D showed in Fig.\ref{res_c} and Fig.\ref{res_d} do not decrease over time.
% It was confirmed that the trust scores of threat models E and F showed in Fig.\ref{res_e} and Fig.\ref{res_f} did not converge to a constant value and oscillated.

%--------------------------------------------------------------------------------------------------------------------
\section{Discussion}\label{discussion}
Past studies have shown that EigenTrust can detect the simple attacks Models A and B but not others in P2P communication (Table~\ref{tab:threat_model}). 

In our study, for Model A, which represents the simplest type of fraud, EigenTrust algorithm properly calculated the attackers’ trust score as being close to 0. 

For Model B, which involves collusion among peers, the algorithm also properly calculated the trust score, which rose sharply at the start of the attack. Although the algorithm did not lower the trust score to 0, it was able to detect attackers by focusing on the decreasing trend. The sharp increase is thought to be due to rates of collective attacks. An attacker who has gone through an incubation period is likely to be selected as a trading partner by normal users due to their high trust score, which rose sharply. As the number of transactions with normal users increases, the trust score gradually decreases as legitimate ratings of bad services accumulate. Ultimately, the attacker’s trust score becomes small enough even if not 0. Therefore, the algorithm can detect malicious activities in C2C services.

For Model C, which includes stochastic service attacks, EigenTrust algorithm failed to detect attacks because the attacker’s trust score did not decrease after the attack started and increased rapidly. Compared with Model B, Model C does not always attack services but instead provides honest services probabilistically in order to deceive normal users. Because users with higher trust scores are more likely to be selected as business partners, the stochastic behaviour may have made detecting dishonest services difficult. In addition, as in Model B, the attacker maintains a high trust score because they assign a high trust score only among allies, and the weight of the attacker’s ratings is important.

For Model D, which includes spies, EigenTrust algorithm again failed to detect the attack; the trust score did not decrease after the attack started but began increasing rapidly. In an extension of Model B, Model D additionally includes spies who not only provide honest services but also conduct rating attacks. Because users with higher trust scores are more likely to be chosen as business partners in the experiment, the spies who provide honest services are likely to be highly evaluated by normal users, which thereby raises the overall trust score of the attackers. In addition, as in Model B, attackers are rated highly only among allies, so the high trust score is maintained, and the weight of each attacker’s evaluation is emphasised.

For Model E, which performs the stochastic rating attack, EigenTrust algorithm calculated the attacker’s trust score as being sharply higher after the start of the attack but oscillating thereafter. Compared with Model C, Model E deceived general users by probabilistically performing a rating attack. The assumption of the experiment is that rating attacks are rated highly only by the attacker’s peers. By contrast, when the attacker evaluates a general user, the evaluation value for the general user is set to 0. Because the algorithm affects the trust score based on the rating that a user receives, rating attacks are likely to heavily impact the trust score in the C2C simulation setting because both buyers and sellers rate each other. In addition, because rating attacks are performed probabilistically, there is a large impact when they are performed and when they are not. Therefore, the trust score is considered to oscillate.

For Model F, which performs the stochastic rating attack, EigenTrust algorithm calculated the attacker’s trust score to be sharply higher after the start of the attack, after which it oscillated as well. Model F, following Model D, deceived general users by performing a rating attack with probability. The trust score oscillated for the same reason as described in Model E.

In addition, while it was previously considered to be unable to detect models C and D, the same results were obtained in this simulation.
On the other hand, the trust scores for models E and F oscillated.
When the trust score oscillates, it can be considered that the user is behaving objectively in an unnatural manner and should be distinguished from a normal user.
If the oscillation of the trust score can be detected, it is considered possible to detect the attacker.
In this simulation, the probability of a rating attack is uniform due to camouflage.
Therefore, one example of a method for detecting trust oscillations is to obtain the distribution of trust scores at a certain time and confirm that the variance is higher than the trust score of a normal user.
Although previous papers discussed trust scores by focusing on the final trust score obtained rather than dynamics, we believe that observing trust scores as a function of time can contribute to the detection of attacks.

Although EigenTrust algorithm was previously considered to be unable to detect Models C and D, the same results were obtained in the simulation. Even so, the trust scores for Models E and F oscillated. When a trust score oscillates, the user is behaving unnaturally objectively and should be distinguished from normal users. If the oscillation of the trust score can be detected, then it is considered to be possible to detect the attacker as well.
In the simulation, the probability of a rating attack is uniform due to camouflage. Therefore, an exemplary method of detecting oscillations in trust is to obtain the distribution of trust scores at a certain time and confirm that the variance is higher than the trust score of a normal user. Although the literature discusses trust scores by focusing on the final trust scores obtained, not their dynamics, we believe that observing trust scores as a function of time can contribute to the detection of attacks.
In relation to the intensity of rating attacks, Model E is an advanced version of Model C, and Model F is an advanced version of Model D. Despite the sophistication of each attack, in some cases the trust score was low due to the oscillation. That outcome indicates that camouflage is counterproductive to rating attacks and that a higher trust score can be maintained if an attack is always performed. As mentioned, Models C and D cannot be detected by EigenTrust because the attackers collude with each other to maintain a high reputation. In that case, aside from devising a method of calculating trust scores, a method of detecting collusion among attackers and eliminating them by clustering networks with high ratings among users can be considered\cite{socialtrust,sybilguard}. 
In addition, because individual subjectivity is important in relation to trust, it is possible to use a trust management algorithm that emphasises either users with transactional tendencies similar to one’s own or users whom one believes in\cite{peertrust,dytrust,socialtrust}. 
In either case, the algorithm is expected to reduce the impact of the reputation of colluding attackers.

%--------------------------------------------------------------------------------------------------------------------
\section{Conclusions}\label{conclusion}
This paper has examined the applicability of EigenTrust algorithm to the use case of C2C services using an agent-based model. The algorithm was developed to calculate the trust scores of nodes in P2P communication and does not take into account the actual situation of C2C services. Therefore, we established transaction protocols for C2C services, in which the buyer and seller rate each other. When analysing the results of the simulations, we focused on changes in the counterparty’s trust score over time. We therefore contribute to the literature (1) by having implemented a simulator that clarifies transaction procedures in purchasing goods and performs transactions between users and (2) by analysing the trust scores resulting from EigenTrust algorithm to confirm the difference in the detection of attacks between P2P and C2C.

We found that EigenTrust algorithm can correctly estimate low trust scores for two types of simple attacks—namely, Models A and B—which is consistent with results in past studies [10]. Therefore, the algorithm is expected to mitigate the impact of simple attackers on C2C services. Furthermore, we found oscillatory patterns in trust scores for two types of advanced attacks—namely, Models E and F—whereas no previous studies have focused on such dynamics. That result suggests that by detecting such oscillations, EigenTrust algorithm may be able to detect advanced attacks that were considered beyond its reach in P2P communications. Doing so was possible because we focused on the transient dynamics of trust scores, whereas past studies have focused on trust scores at a fixed point at the end state of transactions.

Our results provide insights for the further development of Trust Management Algorithms for the benefit of consumer services. How to efficiently detect the oscillation of trust scores to prevent fraud before it happens requires further investigation, however. Other research could better incorporate users’ reputations to improve the algorithm’s ability to detect advanced frauds, including Models C and D. Those improvements may help to increase the trustworthiness of transactions in C2C services.

\newpage
%\bibliographystyle{junsrt} %

\section{Competing interests}
The authors declare no competing interests.

\section{Data availability}
The data set used in this study is available from the corresponding author at a reasonable request. 

\section{Ethical approval}
This study does not need ethical approval statements. 
All data is generated.

\section{Informed consent}
This research did not require informed consent as it is a review paper.

\end{document}